\documentclass{aa}
\usepackage{txfonts}
\usepackage{graphicx}
\usepackage{natbib}
\usepackage{amssymb}
\usepackage{amsfonts}
\usepackage{amsbsy}
\usepackage{amsmath}
\usepackage{pdflscape}

\bibpunct{(}{)}{;}{a}{}{,}

\newcommand{\PFA}{\rm PFA}
\newcommand{\SPFA}{\rm SPFA}
\newcommand{\PD}{\rm PD}
\newcommand{\Ex}{\rm E}

\newcommand{\gb}{\boldsymbol{g}}
\newcommand{\nb}{\boldsymbol{n}}
\newcommand{\ssb}{\boldsymbol{s}}
\newcommand{\xb}{\boldsymbol{x}}
\newcommand{\Xb}{\boldsymbol{X}}

\newcommand{\oneb}{\boldsymbol{1}}

\newcommand{\Hc}{\mathcal{H}}
\newcommand{\Pmatc}{\mathcal{P}}

\newcommand{\ft}{\hat{f}}
\newcommand{\Ft}{\hat{F}}
\newcommand{\st}{\hat{s}}
\newcommand{\ut}{\hat{u}}

\newcommand{\wt}{\hat{w}}

\newcommand{\fmatf}{\mathfrak{f}}
\newcommand{\fmatfb}{\boldsymbol{{\mathfrak f}}}

\begin{document}
   \title{Matched filter in low-number count Poisson noise regime: \\ an efficient and effective implementation}

   \author{R. Vio\inst{1}
    \and P. Andreani\inst{2}
   }
   \institute{Chip Computers Consulting s.r.l., Viale Don L.~Sturzo 82,
              S.Liberale di Marcon, 30020 Venice, Italy\\
              \email{robertovio@tin.it},
          \and
 		  ESO, Karl Schwarzschild strasse 2, 85748 Garching, Germany\\
		  \email{pandrean@eso.org}
             }

\date{Received .............; accepted ................}

\abstract{The matched filter (MF) is widely used to detect signals hidden within the noise. If the noise is Gaussian, its performances are well-known and 
describable in an elegant analytical form. The treatment of non-Gaussian noises is often cumbersome as in most cases there is no
analytical framework.  This is true also for Poisson noise which, especially in the low-number count regime, presents the additional difficulty to be discrete. 
For this reason in the past methods have been proposed based on heuristic or semi-heuristic arguments. Recently, an analytical form of the MF has been introduced but the computation of the probability of false detection or false alarm  ($\PFA$)  
is based on numerical simulations. To overcome this inefficient and time consuming approach we propose here an effective method to compute the $\PFA$  based on the saddle point approximation (SA).
We provide the theoretical framework and support our findings by means of numerical simulations. We discuss also the limitations of the MF in practical applications.}

\titlerunning{Matched Filter with Poisson Noise}
\authorrunning{R. Vio \& P. Andreani}
\maketitle

\section{Introduction}

Detecting signals embedded in noise is a challenge in many research and engineering areas. In the common case of Gaussian noise the standard tool is the matched filter (MF)  \citep{tuz01, min02, osc02, bri04,  lev08, yao13}. According to the Neyman-Pearson theorem, the MF provides the greatest probability of true detection under the condition of a constant probability of false detection  \citep[e.g.][]{kay98}. In non-Gaussian cases it is difficult to derive an MF and even if available it is more cumbersome to use. In particular, it is complex to compute the probability of false detection or false alarm (PFA). In astronomy this situation occurs, for example, in the search for high-energy point sources in the presence of Poisson (i.e. non-additive) noise and where signal and noise together consist of only a few counts per pixel. For this reason, in the past alternative procedures based on heuristic or semi-heuristic arguments have been preferred \citep[][ and reference therein]{ste06}. In a recent work \citet{ofe17}   provide an analytical form  of the MF with the PFA computed from numerical simulations. This approach lacks flexibility and is time-consuming.
In the present paper we propose a method to compute the PFA based on the saddle-point approximation (SA) which is fast, flexible and provides accurate results.

In Section~\ref{sec:formalization} the problem of the detection of a signal in Poisson noise is formalized whereas the SA method is described in Sect.~\ref{sec:spa} and its application illustrated in
Sect.~\ref{sec:spapoiss}. The results of a few numerical experiments are given in Sects.~\ref{sec:numerical}-\ref{sec:psf} and the limitations of the MF in practical situations are discussed in Sect.~\ref{sec:practical}. Finally, the conclusions are given in Sect.~\ref{sec:conclusions}.

\section{The mathematics of the problem}  \label{sec:formalization}

In this section we describe the basic properties of the MF in the case of Poisson noise. 
To allow a better understanding we develop the main arguments in one dimension. The extension to higher dimension cases is straightforward and can be done substituting the coordinate system with a multi-dimensional one
 \footnote{In the case of a two-dimensional map $\Xb$, a one-dimensional signal $\xb$ is obtained through $\xb = {\rm VEC}[\Xb]$, with ${\rm VEC}[.]$ the operator that transforms a matrix into a column array by stacking its columns one underneath the other.}.

To illustrate the procedure of detection of a deterministic and discrete signal of known structure $\ssb = [s(0), s(1), \ldots, s(N-1)]^T$,  with length $N$, and the symbol $^T$ denoting a vector or matrix transpose, we assume the following:
\begin{itemize}
\item The searched signal takes the form $\ssb = a \gb$ with $a$ a positive scalar quantity (amplitude) and $\gb$ typically a smooth function often somehow normalized (e.g., $\sum_{i=0}^{N-1} g(i)  = 1$); \\
\item The signal  $\ssb$ is contaminated by a Poisson noise, i.e. the observed signal $\xb$ is given by $\xb = \Pmatc(\ssb + \lambda \oneb )$. Here, $\oneb =  [1, 1, \ldots, 1]^T$, the scalar $\lambda$ represents the intensity parameter of the noise background and $\Pmatc(\mu \oneb)$ denotes a Poisson random vector with independent entries and expected value (i.e. mean)  ${\rm E}[\xb]=\mu$.
Although not necessary in what follows, this implies that $\lambda$  is constant across $\xb$.
\end{itemize}   
Under these conditions, the detection problem consists of deciding whether $\xb = \nb =  \Pmatc(\lambda \oneb)$, i.e. it is pure noise (hypothesis $H_0$), or it does contain a contribution from a signal $\ssb$ (hypothesis $H_1$). In this way, it is equivalent to a decision problem between the two hypotheses
\begin{equation} \label{eq:decision}
\left\{
\begin{array}{ll}
\Hc_0: & \quad \xb = \Pmatc(\lambda \oneb); \\
\Hc_1: & \quad \xb = \Pmatc(\ssb + \lambda \oneb).
\end{array}
\right.
\end{equation}

Under $\Hc_0$ the probability density function (PDF) of $\xb$ is given by $p(\xb| \Hc_0)$ whereas under $\Hc_1$ by $p(\xb| \Hc_1)$.
Deciding between these two alternatives requires fixing the detection criterion.
A common and effective criterion consists in maximizing the probability of detection ($\PD$)
under the constraint that the $\PFA$ (i.e., the probability of a false detection) does not exceed a fixed value
$\alpha$. The Neyman-Pearson theorem \citep[e.g., see ][]{kay98} allows designing
a decision process that pursues this aim:
to maximize $\PD$ for a given $\PFA=\alpha$, choose $\Hc_1$ 
if the likelihood ratio 
\begin{equation} \label{eq:ratio}
L(\xb) = \frac{p(\xb| \Hc_1)}{p(\xb| \Hc_0)} > \gamma,
\end{equation}
where the threshold $\gamma$ is found from
\begin{equation} \label{eq:p1}
\PFA = \int_{\{\xb: L(\xb) > \gamma\}} p(\xb| \Hc_0) d\xb = \alpha.
\end{equation}

In a recent work \citet{ofe17} show that criteria~\eqref{eq:ratio} and \eqref{eq:p1} lead to the test
\begin{equation} \label{eq:T}
T(\xb) = \xb^T \fmatfb > \gamma,
\end{equation}
where
\begin{equation} \label{eq:mf}
 \fmatfb=\ln{\left(\oneb + \frac{\ssb}{\lambda}\right)}.
\end{equation}
In practice, the test consists of filtering the signal $\xb$ with $\fmatfb$ (i.e. the matched filter) and checking if the statistic $T(\xb)$ exceeds the threshold $\gamma$. From Eq.~\eqref{eq:mf} it appears 
that the form of $\fmatfb$ depends on the signal $\ssb$. 
The issue here is  that in many practical applications (e.g. detection of point sources in sky maps) only the template $\gb$ of the signal is known but not its amplitude $a$. The consequence of using the MF with an incorrect amplitude $a$ is to reduce, for a fixed value of the $\PFA$, the $\PD$, i.e. to make the MF less efficient. \citet{ofe17} have shown that the results provided by the MF are not very sensitive to the precise value of $a$ (see also below). 
This is not surprising given the logarithmic dependence of $\fmatfb$ on $\lambda$.
However, even for a specific value of $a$, the PDF of $T(\xb)$ under the hypothesis $\Hc_0$ is not available in analytical  form. 
The reason is that, although $T(\xb)$ is given by a linear composition of Poisson random variables,  its PDF is not Poissonian. This does not allow fixing the threshold 
$\gamma$ corresponding to a prefixed value $\alpha$ of the $\PFA$. 
\citet{ofe17}  bypass this problem using numerical simulations. Such method, however,  is not flexible and is time-consuming.

An alternative to numerical simulations is approximating the unknown PDF of $T(\xb)$  by another PDF. For example, in the situations of high-number count noise regimes, the Gaussian PDF would be a good choice.
The same does not hold in low-number count regimes. In this case the SA represents an effective solution.

\section{Saddlepoint approximation basics} \label{sec:spa}

The SA is a powerful tool able to provide accurate expressions of the PDFs and the corresponding cumulative distribution functions (CDF). Its derivation is rather technical. For this reason, 
we provide here only a basic introduction useful for practical applications.  A simple and informal derivation is outlined in appendix~\ref{sec:appB}, whereas a rigorous derivation is given in \citet{but07}.

The SA to the PDF $f_X(x)$ of a continuous random variable $X$ is given  by
\begin{equation} \label{eq:aPDF}
\ft_X(x) = \frac{1}{\sqrt{2 \pi K_X^{(2)}(\st)}} \exp{(K_X(\st) - \st x)},
\end{equation}
where $K_X(s)$ is the cumulant generating function (CGF) of $f_X(x)$,
\begin{equation}
K_X(s) = \ln{ M_X(s)} ,
\end{equation}  
and $M_X(s)$ the corresponding moment generating function (MGF) (see appendix~\ref{sec:appA}). Moreover, 
$\st$ is the unique solution to the equation
\begin{equation} \label{eq:K1}
K_X^{(1)}(\st) = x,
\end{equation}
with $K_X^{(j)}$ denoting the $j$th derivative with respect to $s$. $\ft_X(x)$ will not, in general, integrate to one, although it will usually not be far off. Therefore, it has to be numerically normalized.

The SA results are particularly useful to approximate PDFs not obtainable in analytical form for which the corresponding CGF is available. This situation is typical of a random variable $X$ given by the sum of a set of
independent random variables $\{ X_i \}$, $i=1,\ldots, N$. Indeed, except special cases (e.g., Gaussian), also when the random variables $\{ X_i \}$ share the same PDF,  it is not possible to obtain $f_X(x)$ in closed form.
As explained in appendix~\ref{sec:appA}, the CGF of a sum of independent random variables is given by the sum of the respective CGFs. Hence, if the CGFs of the random variables $\{ X_i \}$  are available, the SA can be applied.

Once that $\ft_X(x) $ is available, the correspoding CDF $\Ft_X(x)$ can be obtained via numerical integration. However, a simple expression, able to provide excellent results, has been proposed by \citet{lug80}
\begin{equation}  \label{eq:aCDF}
\Ft_X(x)=
\begin{cases}
\Phi(\wt)  + \phi(\wt) \left(1/\wt - 1/\ut \right) &  \text{if } x \neq \mu, \\
\frac{1}{2} +  \frac{K_X^{(3)}(0)}{6 \sqrt{2 \pi} K_X^{(2)}(0)^{3/2}}  &  \text{if } x = \mu.
\end{cases}
\end{equation}
Here,  $\phi(.)$ and $\Phi(.)$ represent the standard Gaussian PDF and CDF, respectively, whereas $\wt$ and $\ut$ are given by
\begin{equation} \label{eq:wt}
\wt={\rm sign}(\st) \sqrt{ 2 [\st x - K_X(\st)]},
\end{equation}
with ${\rm sign}(y)$ providing the sign of $y$, and
\begin{equation}
\ut=\st \sqrt{ K_X^{(2)}(\st)}.
\end{equation}
The only difficulty in using $\Ft_X(x)$ concerns its evaluation when $x$ is close to the expected value $\Ex[X]=\mu$  of $X$. In this case the computation of $\Ft_X(x)$  is tricky since $K_X^{(1)}(0)={\rm E}[X]$ and then  the solution of Eq.~\eqref{eq:K1} becomes $\st = 0$. As a consequence, it happens that $K_X(0)=0$ and accordingly,
since $\wt = 0$ from Eq.~\eqref{eq:wt}, the first equation in the system~\eqref{eq:aCDF} becomes useless. This is the reason for the second equation in the system~\eqref{eq:wt}. For practical use, 
however, it is numerically more advantageous to use a
linear interpolation based on the SA to ${\rm E}[X] \pm \epsilon$, where $\epsilon$ is chosen small enough to ensure high accuracy, but large enough to ensure numerical stability.

In the case $X$ is a discrete random variable, the same equations hold with $x$ substituted by $k$, which takes values in the set of the integer numbers, and keeping in mind that $1- \Ft_X(k)$ provides the probability that $X \ge k$.
Although the expressions~\eqref{eq:aPDF} and \eqref{eq:aCDF}  are computable for any value of $x$ whether real or integer-valued, $\ft_X(k)$ and $\Ft_X(k)$ are meaningful approximations of $f_X(k)$ and $F_X(k)$ 
only for integer-valued arguments.

\section{Signal detection in the Poisson noise regime}  \label{sec:spapoiss}

The SA to the PDF and the CDF  of $T(\xb)$ in the Poisson case can be easily computed. Indeed,
under the hypothesis $\Hc_0$, $T(\xb)$  is given by a linear combination (i.e. a sum) of Poisson random variables $X_i$ with common parameter $\lambda$,
\begin{equation} \label{eq:Tx}
T(\xb) = y = \sum_{i=0}^{N-1} \fmatf_i x_i.
\end{equation} 
Hence, its CGF $K_Y(s)$ is given by 
\begin{align}
K_Y(s) &= \sum_{i=0}^{N-1} K_{X_i}(s)  \\
            &= \lambda \sum_{i=0}^{N-1}  ({\rm e}^{\fmatf_i s} -1),
\end{align}
with $s \in (-\infty, + \infty)$. It is elementary to see that $K_Y^{(j)}(s)$ is given by
\begin{equation} \label{eq:seq}
K_Y^{(j)}(s) = \lambda \sum_{i=0}^{N-1} \fmatf_i^j {\rm e}^{\fmatf_i s}.
\end{equation}  

In order to be used in Eqs.~\eqref{eq:aPDF} and \eqref{eq:aCDF}, these functions have to be computed for $s=\st$. This step requires the numerical solution of Eq.~\eqref{eq:K1} that, however, does not present particular difficulties
given that, as explained in appendix~\ref{sec:appB}, $K_X^{(1)}(s)$ is an increasing function for $s \in (-\infty, + \infty)$. 

It is important to stress that, since  Eq.~\eqref{eq:K1} cannot be solved when $x=0$, $\ft_X(x)$ is defined only for $x > 0$. But it is easy to see that for $x=0$ it is $\ft_X(0) = \exp{(- \lambda N)}$. This last one is an exact result not only an approximation.

When using the SA in the present context, it is necessary to consider that the PDF of $T(\xb)$ is of discrete type but it is not defined on a lattice (i.e. $y$ does not take values on a regular grid of numbers). The point is that the arguments presented above hold only for integer values of $k$ or, with minor modifications, when $k$ takes values on a regular grid of numbers. However, except for extremely small values of $\lambda$, this PDF can be considered ``almost continuous''.
Indeed,  in the case of two-dimensional maps, from combinatorial arguments it can be realized that, already with values of $\lambda$ as small as $0.01$ and $\fmatfb$ given by a circular Gaussian with standard deviation $\sigma$ of only $2$ pixels, 
the number of different values that $y$ can take is 
of the order of several thousands.  In any case, there is considerable empirical evidence that the SA is useful and maintains most of its accuracy even with discrete PDFs not defined on a lattice \citep[cf. pag. 27 in ][]{but07}. The numerical simulations presented below confirm this result.

\section{Numerical experiments} \label{sec:numerical}

\begin{landscape}
\begin{figure}
        \resizebox{\hsize}{!}{\includegraphics{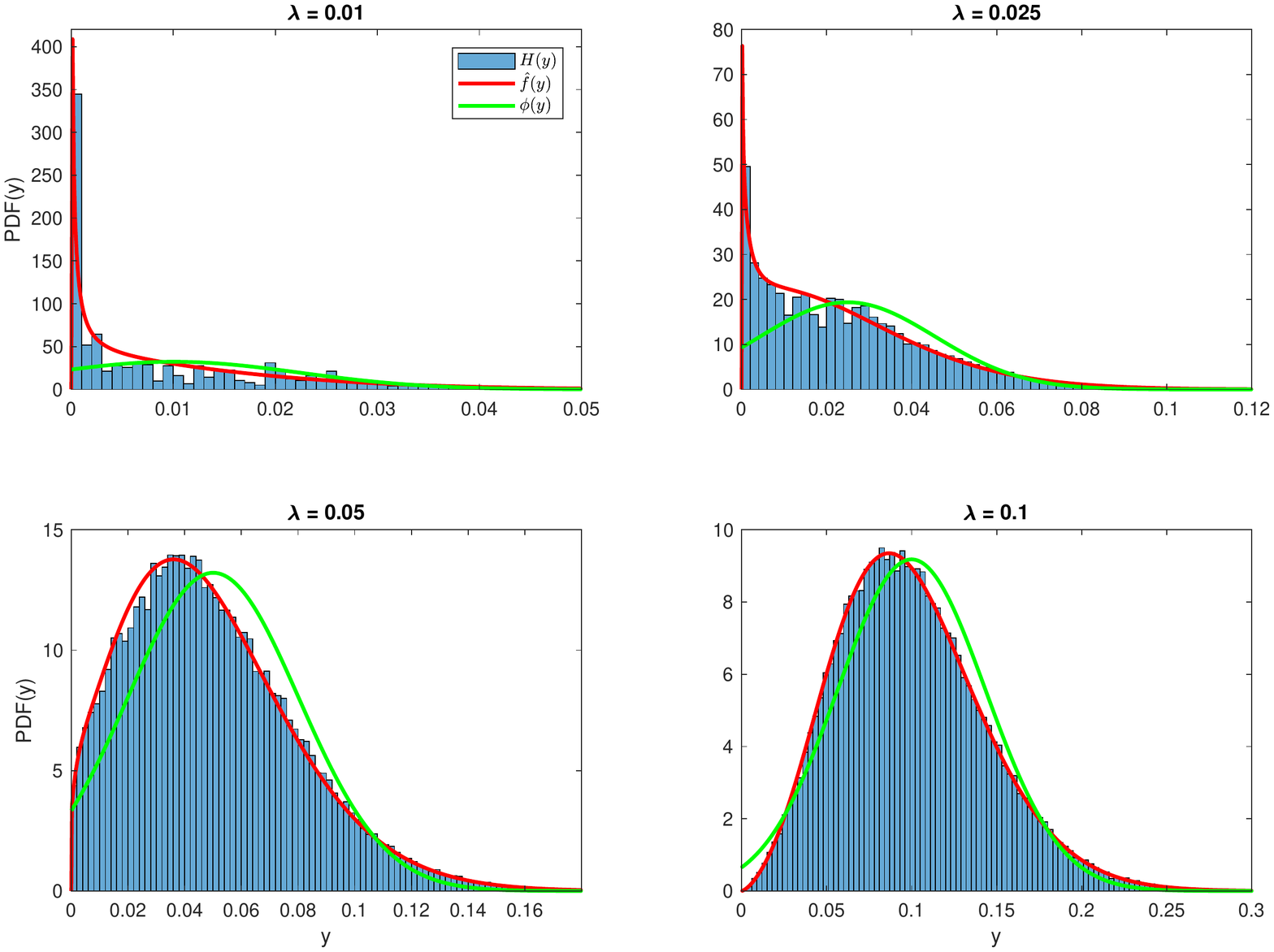}}
        \caption{PDF in the case of the SA (red curve) and the best Gaussian fit (green curve) of the PDF of the statistic $y = T(\xb)$ vs. the corresponding histogram $H(y)$  obtained with a set of $10^5$ MF filtered 
random realizations of a Poisson $13 \times 13$ pixels noise process with the parameter $\lambda$ set to $0.01$, $0.025$, $0.05$, and  $0.1$ counts pix$^{-1}$ (see Sect.\ref{sec:numerical} for details). The used MF is given by Eq.~\eqref{eq:mf} with 
$\ssb$ a circular Gaussian with standard deviation $\sigma=2$ pixels normalized to unit volume.}
        \label{fig:fig_pdf}
\end{figure}
\end{landscape}

\begin{landscape}
\begin{figure}
        \resizebox{\hsize}{!}{\includegraphics{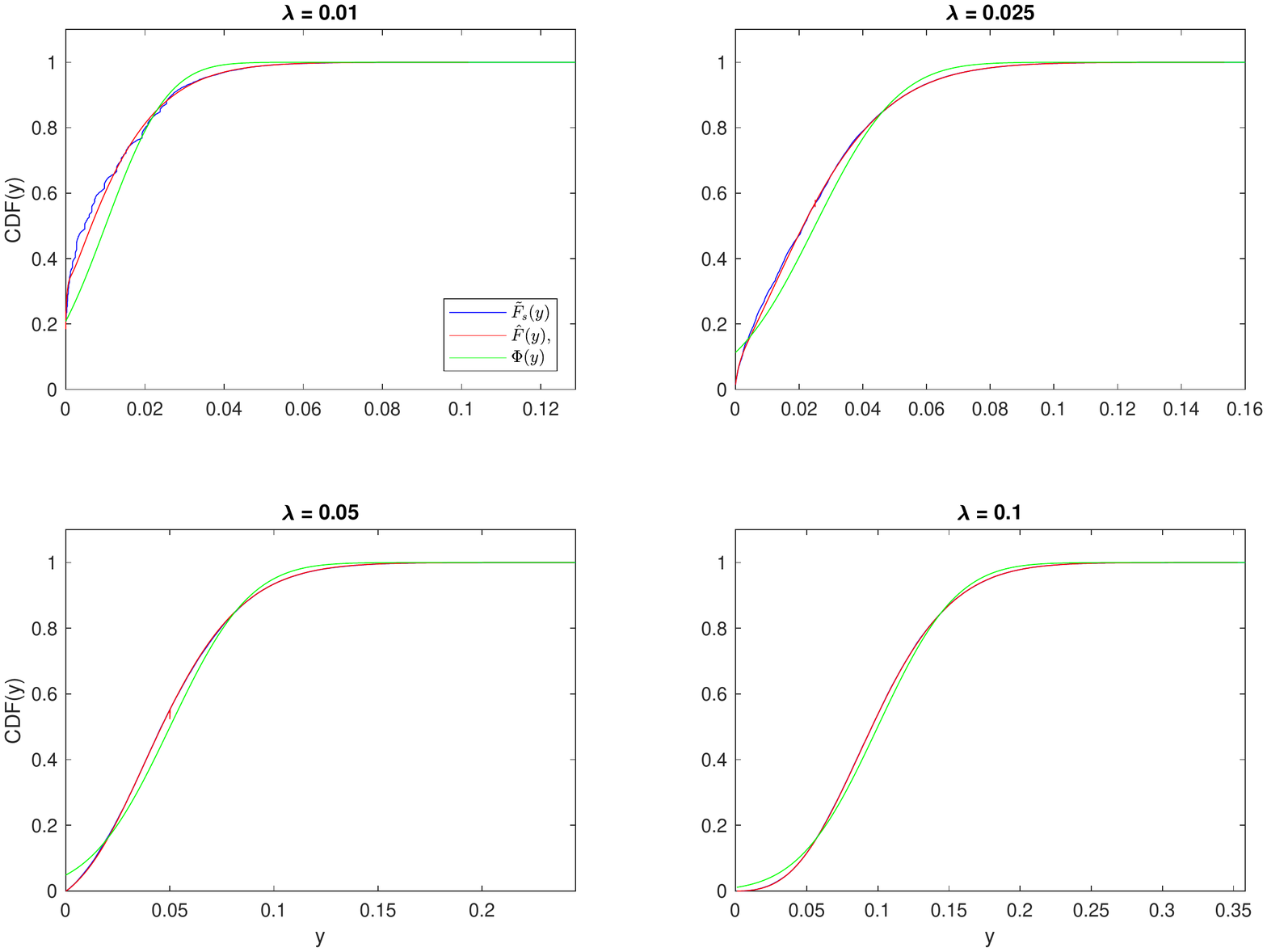}}
        \caption{CDFs corresponding to the the PDFs shown in Fig. 1. The SA is drawn with a red line, the best Gaussian fit with a green line, and the sample CDF with a blue line. In the bottom panels the blue line is not visible since it coincides with the
red one.}
        \label{fig:fig_cdf}
\end{figure}
\end{landscape}

\begin{landscape}
\begin{figure}
        \resizebox{\hsize}{!}{\includegraphics{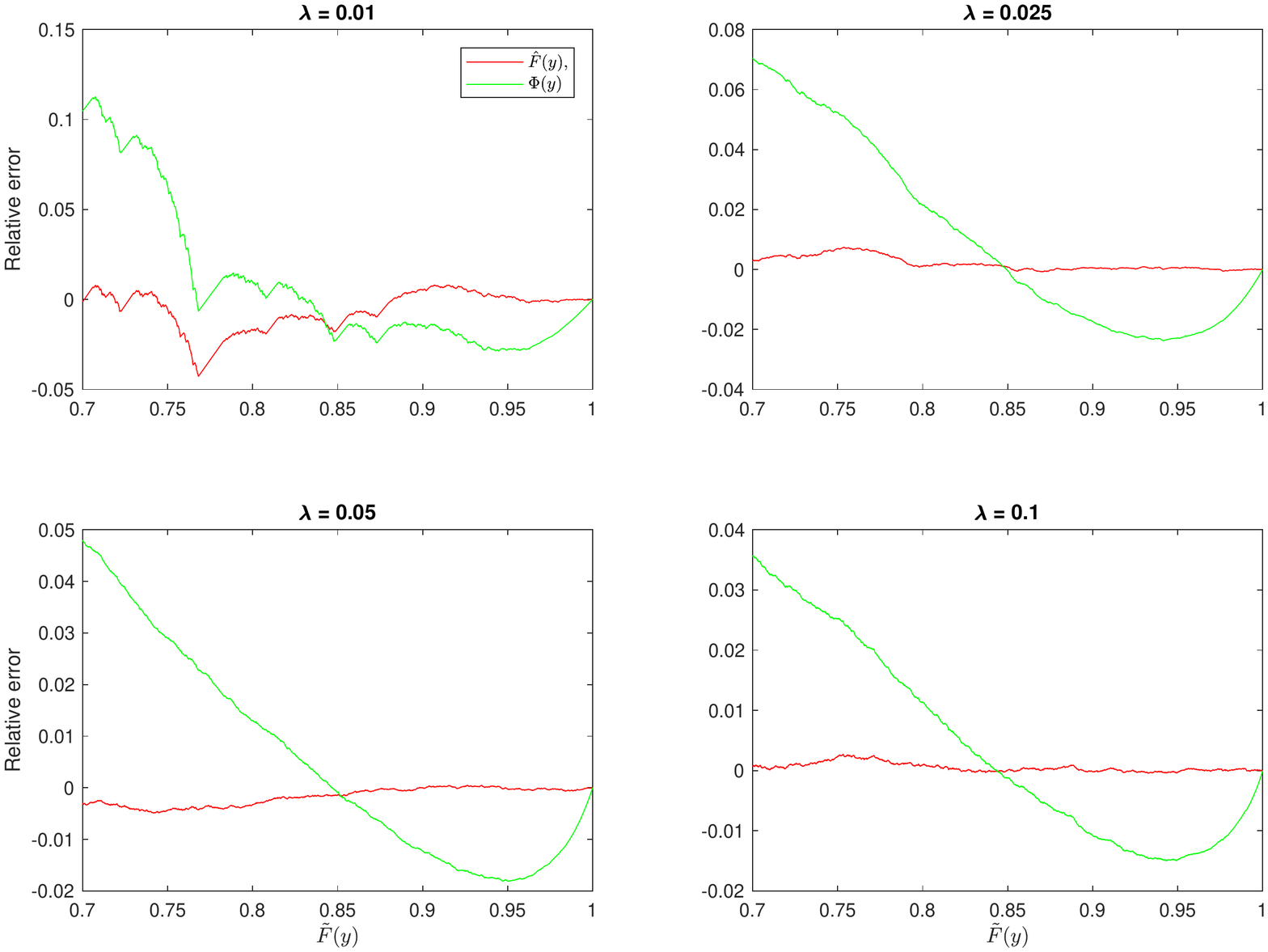}}
        \caption{Relative errors $(\Ft_Y(y)-\tilde{F}_Y(y))/\tilde{F}_Y(y)$  (red line) and $(\Phi(y)-\tilde{F}_Y(y))/\tilde{F}_Y(y)$ (green line) against $\tilde{F}_Y(y)$  for the CDF's in Fig.~\ref{fig:fig_cdf}. The very small relative errors of $\Ft_Y(y)$ highlight its good performance, However, caution has to be used when considering the very right end of these diagrams. This is because the sample CDF $\tilde{F}_Y(y)$ is not an accurate estimate of $F_Y(y)$ when this last is very close to one. Indeed, $F_Y(y) \rightarrow 1$ only when $y \rightarrow \infty$ but $\tilde{F}_Y(y_{\max})=1$ with $y_{\max}$ the greatest among the simulated random variable $Y$.}
        \label{fig:fig_err}
\end{figure}
\end{landscape}

Figure~\ref{fig:fig_pdf} compares the histogram $H(y)$ with the SA $\ft_Y(y)$  (normalized to unit area)  to the PDF $f_Y(y)$ of the statistic $T(\xb)$ for the central pixel of a set of $10^5$ MF filtered random realizations of a 
Poisson $13 \times 13$ pixels noise process. 
Four values of the parameter $\lambda$, say $0.01$, $0.025$, $0.05$, and $0.1$ (units in counts pix$^{-1}$), have been considered. The MF has been computed assuming $\ssb$ as a circular Gaussian with standard deviation $\sigma=2$ pixels 
normalized to unit volume. For comparison, the Gaussian best fits $\phi(y)$ are also shown. From these figures it is visible that better results are obtainable when increasing the values of $\lambda$. This is not an unexpected result since 
small values of $\lambda$ imply that most of the pixels have zero counts, a few  have one count and very few have larger counts.  Especially for ``narrow'' signals $\ssb$, the consequence is a rough PDF for the statistic $T(\xb)$. 
However, in the case of $\lambda=0.01$, the top-right panel of Fig.~\ref{fig:fig_pdf} shows that, although the SA is not able to reproduce all the details of $H(y)$, 
it provides a good envelope resulting in a good approximation to the corresponding CDF. This is supported by Fig.~\ref{fig:fig_cdf} which compares the sample CDF $\tilde{F}_Y(y)$  with the $\Ft_Y(y)$ corresponding to the 
PDFs in  Fig.~\ref{fig:fig_pdf}. The agreement is excellent. This is more evident
in Fig.~\ref{fig:fig_err} where the relative errors $(\Ft_Y(y)-\tilde{F}_Y(y))/\tilde{F}_Y(y)$  are plotted versus $\tilde{F}_Y(y)$ for values of $y$ such as  $\tilde{F}_Y(y) > 0.7$.
This is a useful fact since in signal detection problems it is the complementary CDF $1-F_Y(y)$ and not the PDF $f_Y(y)$ which matters. 

The SA does not work with very small  values of $\lambda$. However, it is questionable that in situations of very low-number count regime the MF could be a useful approach.  For example, in the case
of a $1000 \times 1000$ pixels map and $\lambda=0.001$, only $N_{\emptyset}=1000$ pixels are expected with values greater than zero, i.e. only one pixel in each area of $30 \times 30$ squared pixels. Under these conditions the 
use of the MF does not make sense since there is nothing to filter out. Moreover, the expected noise background should consist of $1000$ bumps all with the same shape and amplitude. A more effective approach is based on the observation that 
the probability to have two counts in a pixel is of order of $5 \times 10^{-7}$. Thereof, all the non-zero pixels are expected to have only one count. 
In addition, given that the probability of two pixels occupying adjacent positions is of order 
of $4 N_{\emptyset}/N_ {\rm pix}$, only four of them are expected to be neighboring. In other words, the detection of a signal can be claimed with high confidence in presence of pixels with counts greater than one and possibly contiguous with other
non-zero pixels.

The conclusion is that, apart from extremely low levels of the noise background, the SA is able to provide excellent results. As visible 
in Figs.~\ref{fig:fig_pdf}-\ref{fig:fig_err}, this is not true for the Gaussian approximation. Hence, to test if after the MF filtering the value $y$ of a pixels is not 
due only  to the noise, it is sufficient to check if $1-\Ft_Y(y) < \alpha$ with $\alpha$ a prefixed $\PFA$.  

\section{Expected performances of the MF} \label{sec:psf}

The excellent computation efficiency, but above all  the good flexibility of the SA in the calculation of the $\PFA$ with different noise levels and functional forms of $\ssb$, allow us to explore the expected performances of the MF in better detail than possible with an approach based on numerical simulations. In particular, in \citet{ofe17} the performances of the MF are compared to those of its most important competitor which is the point spread function filter  (PSFF) technique where it is assumed that $\fmatfb=\gb$.  One of the main reasons for such a comparison lies in the fact that in the case of a Gaussian additive white-noise the PSFF and the MF concide. Another reason is that even for relatively small values of 
$\lambda$ the PDF of the Poisson noise can be reasonably approximated with a Gaussian. Moreover, under the conditions we are working under, the Poissonian noise level is constant across the background area but it changes only close to the signal position. However, since the MF is often used in situations where the amplitude of the signal is smaller than the level of the noise,  in first approximation it can be assumed that $\ssb + \lambda \approx \lambda$. Hence, the noise can be considered of additive type with a constant level everywhere. In conclusion, there are situations where the PSFF $\gb$ can be an acceptable approximation to the MF. 
The results obtained by \citet{ofe17} indicate that in general the MF is superior to the PSFF for small values of $\lambda$ and similar
for larger values of this parameter. But, because of the very large number of numerical simulations necessary to fix a reliable detection threshold for each different experimental situation, their analysis is based on a few cases only. Here, we carry out a set of similar simulations with a larger combination of noise levels and signal intensities.

\begin{landscape}
\begin{figure}
        \resizebox{\hsize}{!}{\includegraphics{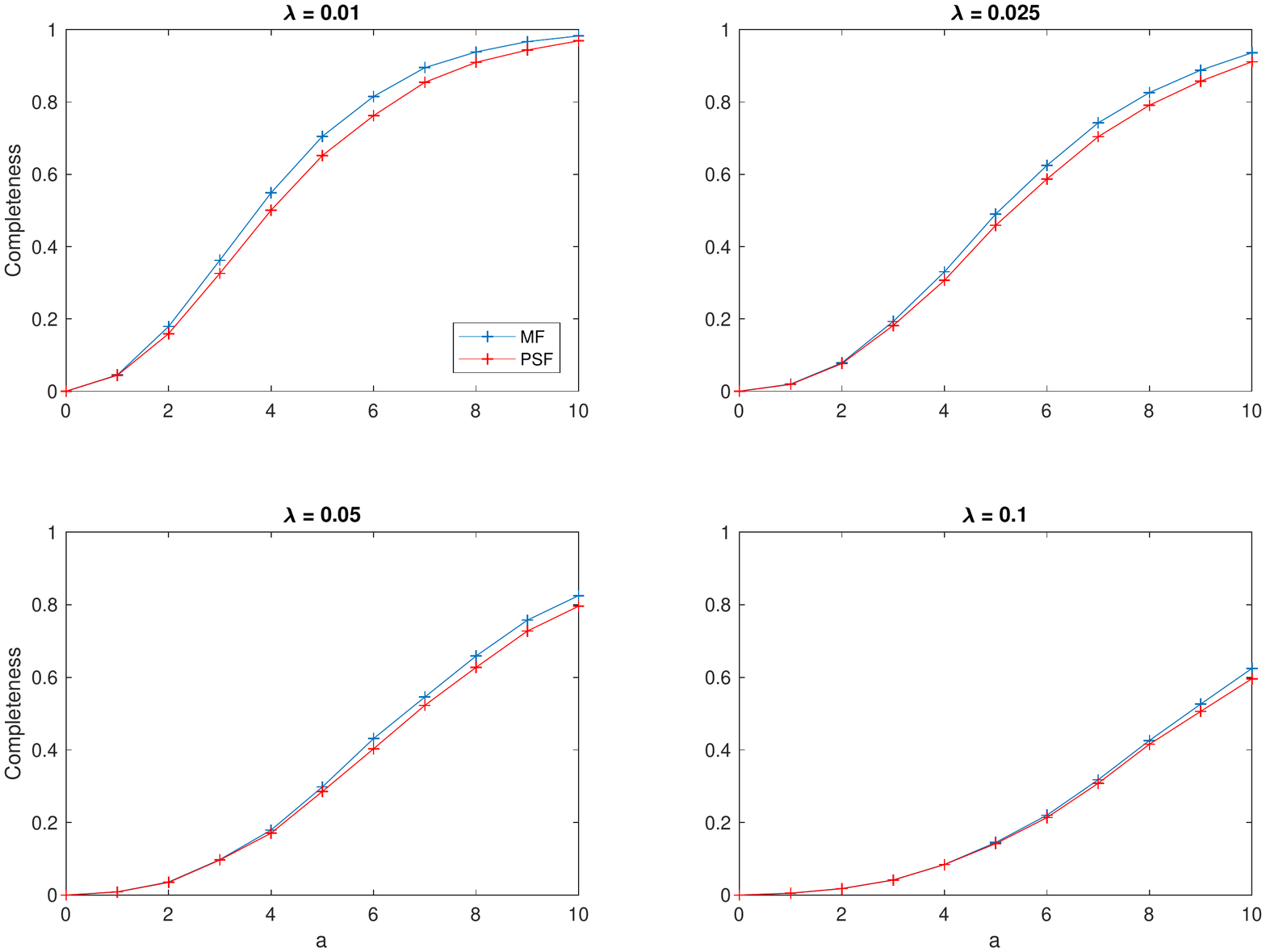}}
        \caption{Results of the numerical experiments presented in Sec.~\ref{sec:psf}. Here, the performances of the MF and of the PSFF are compared on the basis of the completeness (i.e. the fraction of correctly detected signals) for different combinations of the amplitude $a$ of the signal $\ssb$ with  the intensity  $\lambda$ (units in counts pix$^{-1}$) of the background (see Sec.~\ref{sec:psf}). Here, $\ssb$ is given by $a \gb$ with $\gb$ a circular
Gaussian with $\sigma=2$ pixels normalized to unit volume. $10^4$ simulated $13 \times 13$ pixels maps have been used. The detection test has been run assuming a detection threshold corresponding to $\alpha = 10^{-3}$.}
        \label{fig:fig_comparison}
\end{figure}
\end{landscape}

Our numerical experiments confirm the results of \citet{ofe17}.  This is shown by Fig.~\ref{fig:fig_comparison} where the completeness (i.e. the fraction of correctly detected signals) of the MF is compared with that of the PSFF for 
various combinations of $\lambda$ and $a$. It is visible as the results provided by the MF are effectively superior to those of the PSFF. However, the performances move closer and closer for increasing values of $\lambda$.
In the experiments $\gb$ is a circular Gaussian with $\sigma = 2$ pixels, and the completeness has been estimated on the basis of $10^4$ simulated $13 \times 13$ pixels maps assuming a detection threshold corresponding to $\alpha = 10^{-3}$.

\begin{landscape}
\begin{figure}
        \resizebox{\hsize}{!}{\includegraphics{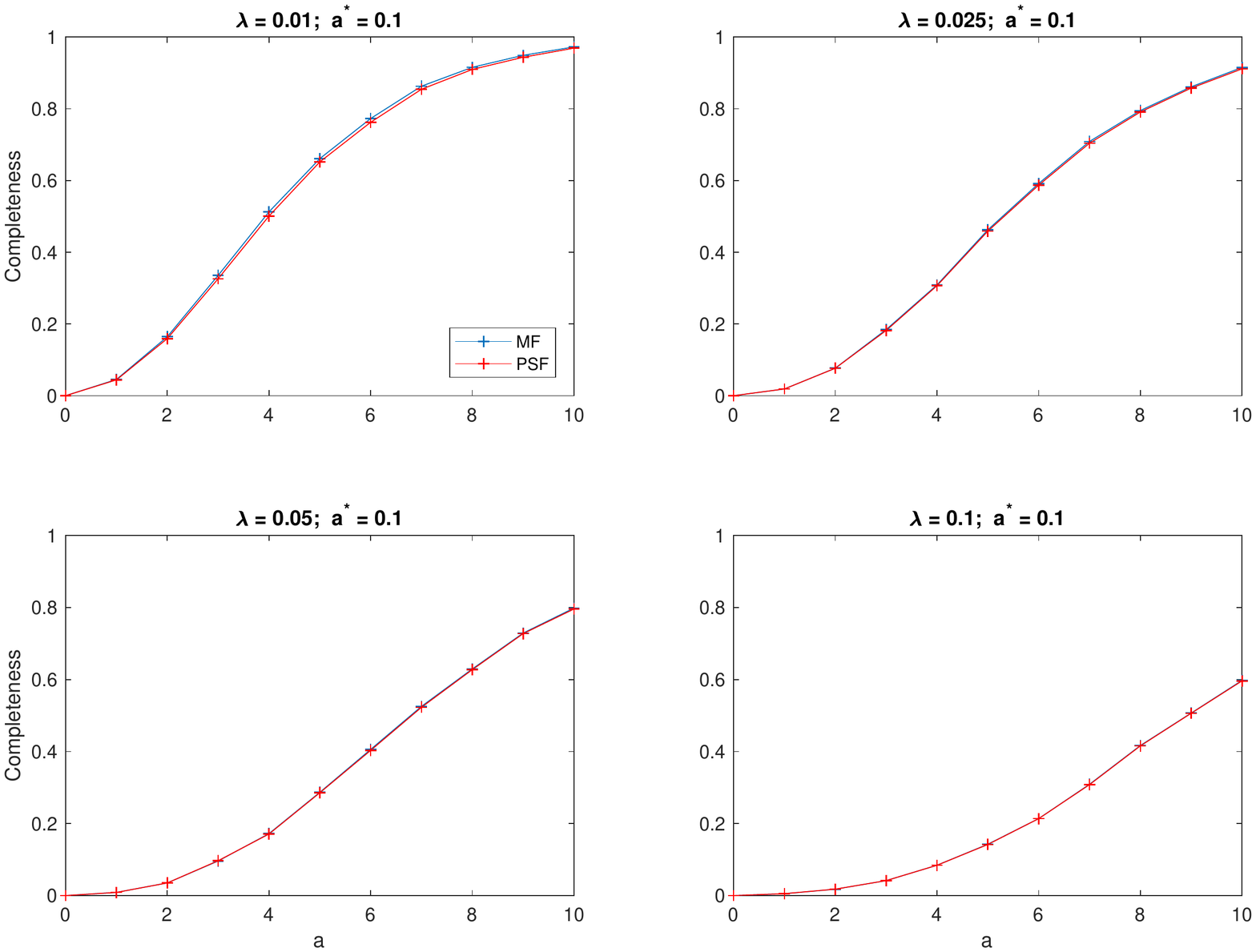}}
        \caption{As in Fig.~\ref{fig:fig_comparison} but the MF has been computed using an incorrect value $a^*=0.1$ for the amplitude $a$ of the signal.}
        \label{fig:fig_comparison_f_01}
\end{figure}
\end{landscape}

\begin{landscape}
\begin{figure}
        \resizebox{\hsize}{!}{\includegraphics{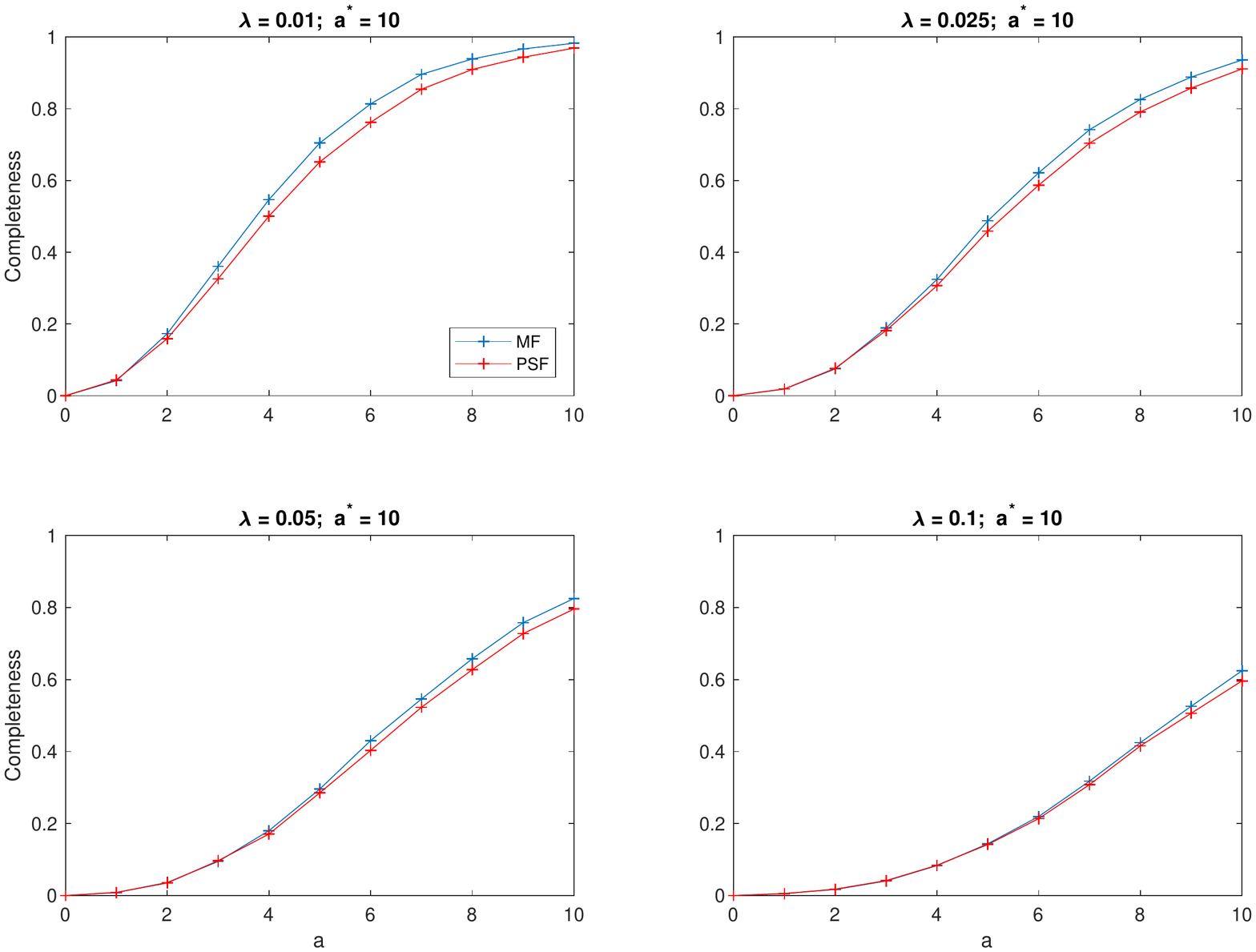}}
        \caption{As in Fig.~\ref{fig:fig_comparison} but the MF has been computed using an incorrect value $a^*=10$ for the amplitude $a$ of the signal.}
        \label{fig:fig_comparison_f_10}
\end{figure}
\end{landscape}

\begin{landscape}
\begin{figure}
        \resizebox{\hsize}{!}{\includegraphics{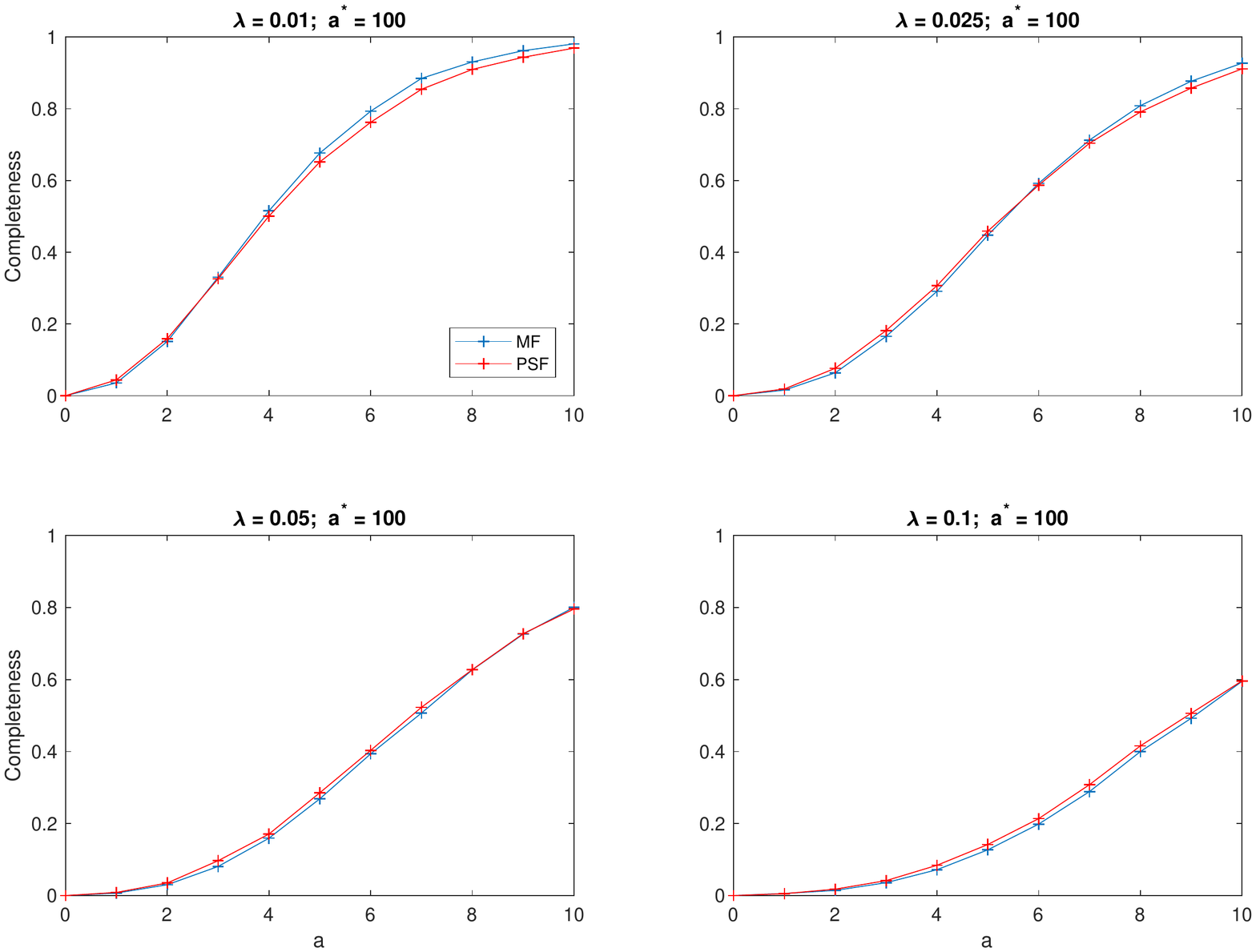}}
        \caption{As in Fig.~\ref{fig:fig_comparison} but the MF has been computed using an incorrect value $a^*=100$ for the amplitude $a$ of the signal.}
        \label{fig:fig_comparison_f_100}
\end{figure}
\end{landscape}

Although indicative, such experiments, as well those by  \citet{ofe17}, are affected by an important limitation: the comparison between the MF and the PSFF is done assuming that $a$ in the MF takes true values. But in real situations this information is not available. An example is sky maps containing point sources with different amplitude. For this reason, we have carried out another set of numerical experiments
similar to the previous ones but using incorrect signal amplitudes $a^*$.  The results are shown in Figs.~\ref{fig:fig_comparison_f_01}-\ref{fig:fig_comparison_f_100} for $a^*=0.1$, $10$, and $100$. 
The indication that comes out from these figures is that, for a given value of $\lambda$,  the performances of the 
MF effectively do not critically depend on the exact value of $a$ and, although less remarkably, tend to remain superior to those provided by the PSFF. Only values of $a^*$ very different from the true one determine an appreciable degradation of the results. 

\begin{landscape}
\begin{figure}
        \resizebox{\hsize}{!}{\includegraphics{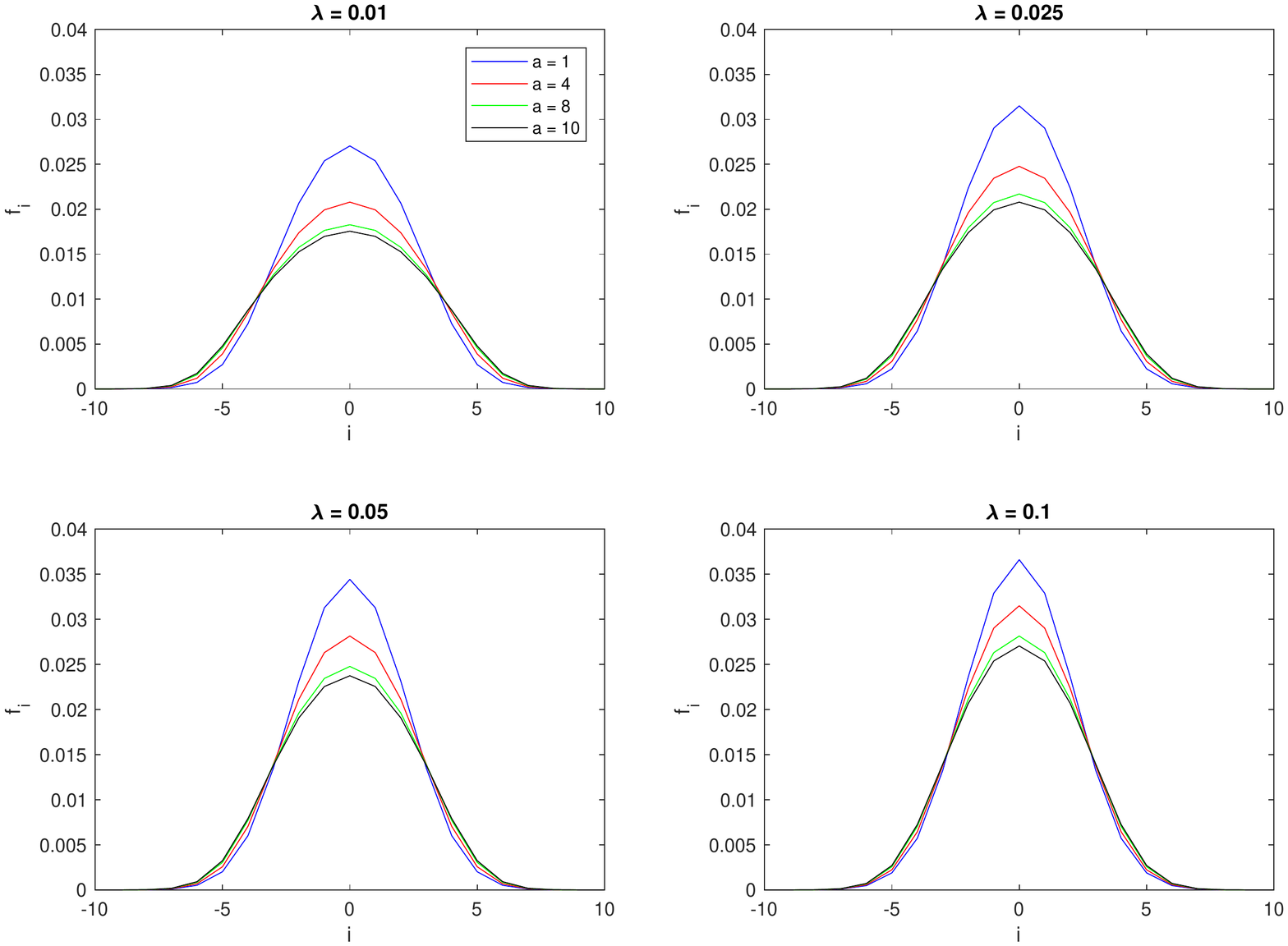}}
        \caption{1-D cuts of the MF for different values of the signal amplitude $a$ and fixed value of the intensity parameter $\lambda$ of the background. }
        \label{fig:fig_filters}
\end{figure}
\end{landscape}

Finally,  another indication which comes out  from Figs.~\ref{fig:fig_comparison_f_01}-\ref{fig:fig_comparison_f_100} is that it appears less harmful to use values of $a^*$ greater, rather than smaller, the true value. The reason of this fact can be deduced from Fig.~\ref{fig:fig_filters}  which shows the 1-D cut of the MF for different combinations of $\lambda$ and $a$. With $\lambda$ fixed, it is evident that when $a$ increases the filtering action of the MF is stronger. This results in an over-filtered signal but also in a more robust attenuation of the noise.  On the contrary, using an $a^*$ smaller that $a$ will result in an insufficiently filtered noise. The former situation appears preferable. 

\section{Shortcomings of the MF with Poisson noise} \label{sec:practical}

\begin{landscape}
\begin{figure}
        \resizebox{\hsize}{!}{\includegraphics{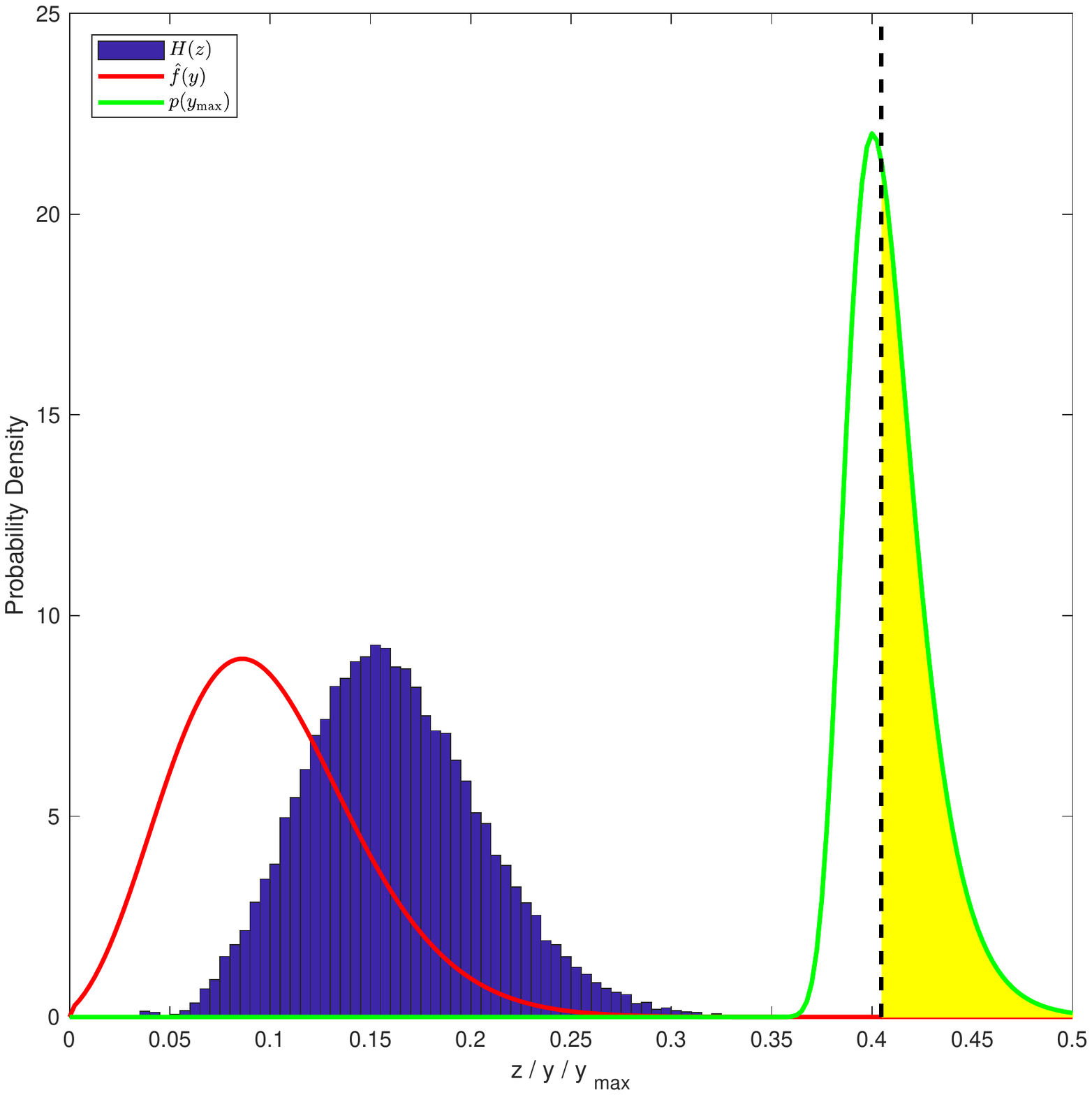}}
        \caption{Comparison of the SA $\ft_Y(y)$ (red line) of the PDF of the counts of a MF filtered $2000 \times 2000$ pixels Poisson random field ($\lambda=0.1$ counts pix$^{-1}$) with the histogram $H(z)$ of its peaks. 
        The used MF is given by Eq.~\eqref{eq:mf}  with $\ssb$ a circular Gaussian with $\sigma=2$ pixels and normalized to unit volume.
        For reference,  the PDF $p_Z(z)$  (green line) of the greatest value of the map is also drawn (see text). 
        The colour-filled area provides the $\SPFA$ for the greatest peak effectively observed in the simulated map ($\SPFA \approx 0.47$).}
        \label{fig:fig_pdf_peaks}
\end{figure}
\end{landscape}

Thanks to the SA it is possible to find a partial solution to a still unsolved problem. In particular, the performances of the 
MF~\eqref{eq:mf} have been tested under the implicit assumption that the position of $\ssb$ within $\xb$ is known. In real situations this condition is rarely met (e.g. point sources in a two-dimensional map).  
The standard procedure to circumvent this issue is to cross-correlate $\xb$  with the MF and then to apply the 
detection test~\eqref{eq:T} to the resulting most prominent peaks.  However, as shown in \citet{vio16}  and \citet{vio17} for the Gaussian case, the PDF 
of the peaks $\{ z \}$ of a random field is different from that of its generic points. Of course, the same  holds also for the Poisson case.  This is shown  by 
Fig.~\ref{fig:fig_pdf_peaks} where the PDF $\ft_Y(y)$ of a $2000 \times 2000$ pixels Poisson random field with $\lambda=0.1$ and filtered by means of the MF in Eq.~\eqref{eq:mf} with $\ssb$ 
a circular Gaussian with $\sigma=2$ pixels, is compared with the histogram $H(z)$ of its peaks.
It is evident that working with the peaks, and assuming the PDF $\ft_Y(y)$ for the the statistic $T(z)$ in Eq.~\eqref{eq:Tx}, may severely underestimate the $\PFA$ with the risk of giving statistical significance to features that
belong to the noise. 

Contrary to the Gaussian noise \citep{che15a, che15b}, the PDF of the peaks is not available for the Poisson noise. Without it, a precise computation of the $\PFA$ is not possible. Moreover, there is the additional difficulty 
that, if $N_p$ peaks are present in a MF filtered map, then a number  $\alpha \times N_p$  among them is expected to exceed, by chance, the prefixed detection threshold. For example, if $N_p = 1000$, then there is a high probability that a detection with a 
nominal $\PFA$ equal to $10^{-3}$ is spurious.
Hence, the true $\PFA$ has to depend on $N_p$ (look-elsewhere effect). The popular way-out to get around this issue is to assume
$f_Y(y)$ as the PDF of the peaks (i.e. the peaks are regarded as generic points of the random field) and then to set  $\PFA = \alpha / N^*$
with $N^*$ the number of independent pixels. If the noise is coloured, as it happens after the filtering operation, 
pixels are correlated with each other. Therefore, $N^*$ typically is smaller than $N$ and has to be estimated. Usually the estimation of $N^*$ is based
on the correlation length of the template $\gb$. The rational is that pixels with a mutual distance wider than the correlation length can be considered independent. For instance,  in the case of a two-dimensional map and $\gb$  
given by a circular Gaussian function with dispersion $\sigma$, \citet{ofe17} suggest that $N^* \approx N/\sigma^2$. This is to point out that such approach provides results which are only correct as an order of magnitude but no alternative is 
available if additional a priori information is missing.

In the present context, such procedures can be efficiently implemented by means of the order statistics, in particular by exploiting  the statistical characteristics of 
the greatest value of a finite sample of 
identical and independently distributed (iid) random variables from a given 
PDF  \citep{hog13}. Under the iid condition, the PDF $\psi_{Q}(q)$ and the CDF $\Psi_Q(q)$ of the greatest value $q = y_{\max}$ among a set of $N^*$ independent pixels with PDF $\ft_Y(y)$ are given by
\begin{equation} \label{eq:gz}
\psi_Q(q) = N^* \left[ \Ft_Y(y) \right]^{N^*-1} \ft_Y(y),
\end{equation}
and
\begin{equation} \label{eq:gz}
\Psi_Q(q) = \left[ \Ft_Y(y) \right]^{N^*},
\end{equation}
respectively.
Hence, by assuming that  the greatest value among a set of pixels coincides with that of a peak, a detection can be claimed when for a given peak it is $1-\Psi_Q(q) < \alpha$. 

This is an alternative way, though equivalent, to threshold the data which does not require the inversion of $\Psi_Q(q)$ to fix the parameter $\gamma$ corresponding to a given $\alpha$. In \citet{vio17} the quantity 
$1-\Psi_Q(q)$ is called specific probability of false alarm (SPFA).
The PDF $\psi_Q(q)$ for the numerical experiment of this section is also shown in Fig.~\ref{fig:fig_pdf_peaks} as well the $\SPFA$ corresponding to the largest peak observed  in the map. 
We stress that in the case of large maps, an alternative  based on numerical simulations is not viable.

\section{Conclusions} \label{sec:conclusions}

In this paper we have introduced an efficient and effective implementation of the matched filter in the case of low-number count Poisson noise regime. We have shown that although the probability distribution and the cumulative distribution functions of the pixel counts after the MF filtering are not available they can be approximated with excellent results using the saddlepoint  approximation method. With such techniques more accurate estimations of the probability of false detection or false alarm are obtained without making use of empirical or numerical methods.

\begin{acknowledgements}
The authors warmly thank Martine Pelzer for her careful reading of the paper and correction of the English editing. 
\end{acknowledgements}

\clearpage
\clearpage
\clearpage
\clearpage
\clearpage
\clearpage
\clearpage
\clearpage

\clearpage
\appendix
\section{A simple derivation of the saddlepoint approximation} \label{sec:appB}

Saddlepoint approximation (SA) is a powerful tool for obtaining accurate expressions of probability density functions and cumulative distribution functions. It can be derived in a number of ways. Most of them, however, are rather technical. Here, we provide
an informal derivation due to \citet{pao07}. A more technical and rigorous derivation can be found in \citet{but07}.

As seen in appendix ~\ref{sec:appA}, if $X$ is random variable with PDF $f_X(x)$, moment generating function (MGF) $M_X(s)$ existing for $s \in U$ with $U$ a neighborhood around zero, and cumulant generating function (CGF) $K_X(s) = \ln M_X(s)$, its mean and variance are given by $K^{(1)}_X(0)$ and $K_X^{(2)}(0)$, respectively. With this premise, the derivation of the SA proceeds according to the following steps:
\begin{enumerate}
\item A new random variable $T_s$ is defined having PDF
\begin{equation} \label{eq:Ts}
f_{T_s}(x; s) = \frac{{\rm e}^{xs} f_X(x)}{M_X(s)},
\end{equation}
for some $s \in U$; \\
\item the MGF and the CGF of $T_s$ are computed, respectively, as
\begin{align}
M_{T_s}(t) & = \frac{M_X(t+s)}{M_X(s)}, \\
K_{T_s}(t) & = K_X(t+s) - K_X(s).
\end{align}
The mean and the variance of $T_s$ are given, respectively, by $K_T^{(1)}(0)=K_X^{(1)}(s)$ and $K_T^{(2)}(0)=K_X^{(2)}(s)$; \\
\item Fixed an arbitrary $s_0 \in U$, $f_{T_s}(x; s)$ is approximated by a Gaussian PDF  $\phi(x; \mu_0, \sigma_0^2)$ with mean $\mu_0$ and variance $\sigma_0^2$. Here, $\mu_0 = K_X^{(1)}(s_0)$ and $\sigma_0^2= K_X^{(2)}(s_0)$; \\
\item From Eq.~\eqref{eq:Ts} $f_X(x)$ can be approximated by means of the PDF $\ft_X(x)$
\begin{align}
\ft_X(x) & = \phi(x; \mu_0, \sigma_0^2) M_X(s_0) {\rm e}^{-s_0 x}, \\
& = \frac{1}{\sqrt{2 \pi \sigma_0^2}} \exp\left\{ - \frac{1}{2 \sigma_0^2} (x - \mu_0)^2 \right\} M_X(s_0) {\rm e}^{-s_0 x};
\end{align} \\
\item The value of $s_0$ has to be determined in such a way that, for a fixed $x$,  $\ft_X(x)$ is the best approximation to $f_X(x)$. In this respect, it is well known that the normal approximation to the distribution of a random variable $X$ is accurate
near the mean of $X$, but degrades in the tails. As a consequence,  it is logical to choose a value $\st$ of  $s_0$ such as $\mu_0=x$, i.e.
\begin{equation} \label{eq:Kx}
K_X^{(1)}(\st)=x;
\end{equation} \\
\item Once $\st$ has been fixed, the normal density approximation $ \phi(x; \mu_0, \sigma_0^2)$ becomes $\phi(x; K_X^{(1)}(\st), K_X^{(2)}(\st))$ 
and the approximation to $f_X(x)$ simplifies to
\begin{equation}
\ft_X(x) = \frac{1}{\sqrt{2 \pi K_X^{(2)}(\st)}} \exp\left\{ K_X(\st) - x \st \right\}, \qquad x = K_X^{(1)}(\st).
\end{equation}
\end{enumerate}
The only critical point in this procedure is that $\st$ has to be evaluated, via the solution of Eq.~\eqref{eq:Kx}, for each $x$ where $\ft_X(x)$ has to be computed. In general, this requires a numerical approach. However,  for a fixed $x$ Eq.~\eqref{eq:Kx} has only one solution $\st$ since
it can be shown that $K_X(s)$ is always a strictly convex function when evaluated over the converge interval $U$ of $M_X(s)$. 
Consequently, the mapping $K^{(1)}(s)$ for $s \in U$ into the support of the PDF $f_X(x)$ is a bijection and thereof $K^{(1)}(s)$ is strictly increasing \citep[see pag. 6 in][]{but07}.
Another question is that  $\ft_X(x)$ will not, in general, integrate to one, although it will usually not be far off. Therefore, it has to be numerically normalized.

\section{The moment and cumulant generating functions} \label{sec:appA}

Given a continuous random variable $X$ with probability density function $f_X(x)$ and a positive number $h$ such that for $-h < s <  h$  the mathematical expectation
\begin{equation}
\Ex[{\rm e}^{sX}]=\int_{-\infty}^{+\infty} {\rm e}^{s x}f_X(x) dx
\end{equation}
exists, then this expectation is called the moment generating function (MGF) of $X$ and is denoted by $M_X(s)$. The largest open interval $U=(a, b)$ around zero where $M_X(s) <  \infty$ for $s \in U$ is referred as the convergence strip of $M_X(s)$. 
In the case $X$ is a discrete random variable with discrete PDF $F_X(k)$, $M_X(s)$ becomes
\begin{equation}
M_X(s)=\sum_{k=-\infty}^{\infty}   {\rm e}^{s k} f_X(k) \qquad s \in U.
\end{equation}

The cumulant generating function (CGF) of $M_X(x)$ is defined as
\begin{equation}
K_X(s) = \ln{ M_X(s)} \qquad s \in U.
\end{equation}
The terms $\kappa_i$ in the series expansion $K_X(t) = \sum_{r=0}^{\infty} k_r t^r / r!$ are referred as the cumulants of $X$ so that the $i$th derivative of $K_X(t)$ evaluated at $t=0$ is $\kappa_i$, i.e.
\begin{equation}
\kappa_i = \left.K^{(i)}_X(t) \right|_{t=0}.
\end{equation}
It is straightforward to show that $\kappa_1 = \mu = \Ex[X]$ and  $\kappa_2 = \sigma^2 = \Ex[X^2 - \mu^2]$, i.e. the first two cumulants provide the mean and the variance of $X$.

One important property of the CGF is that in the case of a sequence of independent random variable $X_1, X_2, \ldots, X_N$ for which $X_i$ has CGF $K_i$ determined over $U_i = (a_i, b_i)$, then the CGF of $X= \sum_{i=1}^N X_i$ is
\begin{equation}
K_X(s) = \sum_{i=1}^N K_{X_i}(s) \qquad s \in (\max a_i, \min b_i).
\end{equation} 
In other words, the CGF of a sum of independent random variable is given by the sum of the respective CGFs.

\end{document}